\documentstyle[12pt]{article}

\newcounter{eq}
\newcounter{sc}

\def\overleftrightarrow#1{\vbox{\ialign{##\crcr
 $\leftrightarrow$\crcr\noalign{\kern-1pt\nointerlineskip}
 $\hfil\displaystyle{#1}\hfil$\crcr}}}

\newlength{\minitwocolumn}
\begin{document}

\begin{flushright}
DPUR/TH/46\\
February, 2016\\
\end{flushright}
\vspace{20pt}

\pagestyle{empty}
\baselineskip15pt

\begin{center}
{\large\bf Induced Gravity from Curvature Density Preserving Diffeomorphisms
\vskip 1mm }

\vspace{20mm}
Ichiro Oda \footnote{E-mail address:\ ioda@phys.u-ryukyu.ac.jp}

\vspace{5mm}
           Department of Physics, Faculty of Science, University of the 
           Ryukyus,\\
           Nishihara, Okinawa 903-0213, Japan.\\

\end{center}


\vspace{5mm}
\begin{abstract}
We construct not only an induced gravity model with the restricted diffeomorphisms, that is, transverse
diffeomorphisms which preserves the curvature density, but also that with the full diffeomorphisms. 
By solving the equations of motion, it turns out that these models produce Einstein's equations 
with a certain Newton's constant in addition to the constraint for the curvature density.
In the limit of the infinite Newton constant, the models give rise to induced gravity.   
Moreover, we discuss cosmological solutions on the basis of the gravitational models at hand.      
\end{abstract}

\newpage
\pagestyle{plain}
\pagenumbering{arabic}


\rm
\section{Introduction}

We have recently watched the revival of a gravitational theory, what is called, $\it{unimodular \ gravity}$ 
\cite{Weinberg}-\cite{Saltas}. 
The original idea of unimodular gravity stems from an observation by Einstein that some equations of general relativity
take simpler expressions when they are described in the unimodular coordinates where the determinant of the metric
tensor takes $-1$, that is, $\det g_{\mu\nu} = -1$ \cite{Einstein}.

It is considered that unimodular gravity is obtained from general relativity by choosing a gauge condition
$\sqrt{-g} = 1$ for diffeomorphisms. The resultant equations of motion are then given by the traceless part
of Einstein equations in addition to the unimodular constraint $\sqrt{-g} = 1$. One of the attractive points
in unimodular gravity is that the cosmological constant appears as an integration constant which can take any values. 
This fact has been previously expected to give some insight in the well-known cosmological constant problem. However, 
it might be difficult to resolve the cosmological constant problem within the framework of unimodular gravity 
if we think that unimodular gravity is nothing but a gauge-fixed theory of general relativity as mentioned above.

Before describing our gravitational models in detail in the next section, let us begin with unimodular gravity
since we find it to be useful to explain our basic idea. Let us start with the conventional Einstein-Hilbert action 
with the cosmological constant \footnote{We follow notation and conventions by Misner et al.'s textbook \cite{MTW}, 
for instance, the flat Minkowski metric $\eta_{\mu\nu} = diag(-, +, +, +)$, the Riemann curvature tensor 
$R^\mu \ _{\nu\alpha\beta} = \partial_\alpha \Gamma^\mu_{\nu\beta} - \partial_\beta \Gamma^\mu_{\nu\alpha} 
+ \Gamma^\mu_{\sigma\alpha} \Gamma^\sigma_{\nu\beta} - \Gamma^\mu_{\sigma\beta} \Gamma^\sigma_{\nu\alpha}$, 
and the Ricci tensor $R_{\mu\nu} = R^\alpha \ _{\mu\alpha\nu}$. The reduced Planck mass is defined as 
$M_p = \sqrt{\frac{c \hbar}{8 \pi G}} = 2.4 \times 10^{18} GeV$ where $G$ is the Newton constant. Moreover, as usual, 
we define the gravitational coupling constant $\kappa$ in terms of $\kappa^2 = 8 \pi G = \frac{1}{M_p^2}$.}
\begin{eqnarray}
S_{GR} &=& \frac{1}{2 \kappa^2} \int d^4 x  \sqrt{-g} \left( R - 2 \Lambda \right) \nonumber\\
&=& \frac{1}{2 \kappa^2} \int d^4 x  \sqrt{-g} R -  2 \Lambda^\prime \int d^4 x  \sqrt{-g},
\label{EH action}
\end{eqnarray}
where $\Lambda$ is the cosmological constant and we have introduced the rescaled cosmological
constant $\Lambda^\prime = \frac{1}{2 \kappa^2} \Lambda$. Based on this action, unimodular gravity can be
understood as follows: In unimodular gravity, we focus on the second term, i.e., the cosmological constant term,
putting $\Lambda = 0$, and try to derive effectively the cosmological constant by taking a gauge condition $\sqrt{-g} = 1$ 
for one of diffeomorphisms. 

In unimodular gravity, therefore, the starting action takes the form
\begin{eqnarray}
S_{UM} = \int d^4 x  \left[ \frac{1}{2 \kappa^2} \sqrt{-g} R 
- \lambda(x) \left( \sqrt{-g} - v_0 \right) \right] + S_m,
\label{UM action}
\end{eqnarray}
where $\lambda(x)$ is the Lagrange multiplier field enforcing the unimodular constraint $\sqrt{-g} = v_0$
with $v_0$ being a constant which is introduced for the generality instead of $1$, and $S_m$ means the matter action. 
Note that this action is not manifestly invariant under the full diffeomorphisms because of the presence of the term 
$\lambda(x) v_0$. 

Variation of the metric tensor yields the Einstein equations:
\begin{eqnarray}
\frac{1}{\kappa^2} G_{\mu\nu} + \lambda(x) g_{\mu\nu} = T_{\mu\nu},
\label{Einstein-eq 1}
\end{eqnarray}
where $G_{\mu\nu} = R_{\mu\nu} - \frac{1}{2} g_{\mu\nu} R$ is the Einstein tensor and $T_{\mu\nu} = - \frac{2}{\sqrt{-g}}
\frac{\delta}{\delta g^{\mu\nu}} S_m$ is the energy-momentum tensor of matter. Then, operating $\nabla^\mu$ on the Einstein equations 
(\ref{Einstein-eq 1}), with the help of the Bianchi identity $\nabla^\mu G_{\mu\nu} = 0$ and the conservation law 
$\nabla^\mu T_{\mu\nu} = 0$, one obtains the equations $\partial_\mu \lambda = 0$, which mean that the Lagrange multiplier
field $\lambda(x)$ is a constant, $\lambda(x) = \lambda_0$. As a result, the dynamics of unimodular gravity is classically
equivalent to that of general relativity with a constant cosmological constant $\lambda_0 \kappa^2$. 

Incidentally, sometimes in unimodular gravity, the traceless Einstein equations are regarded as the fundamental equations
\begin{eqnarray}
\frac{1}{\kappa^2} \left( R_{\mu\nu} - \frac{1}{4} g_{\mu\nu} R \right) 
= T_{\mu\nu} - \frac{1}{4} g_{\mu\nu} T,
\label{Traceless Einstein-eq 1}
\end{eqnarray}
which can be obtained as follows: First, taking the trace of the Einstein equations (\ref{Einstein-eq 1}), we have
\begin{eqnarray}
\lambda = \frac{1}{4} \left( \frac{1}{\kappa^2} R + T \right).
\label{lambda-eq 1}
\end{eqnarray}
Next, substituting this expression for the Lagrange multiplier field into the Einstein equations (\ref{Einstein-eq 1}),
it is easy to show that we can get Eqs. (\ref{Traceless Einstein-eq 1}).  

One of the most attractive aspects in unimodular gravity is that the cosmological constant arises as an arbitrary constant
of integration from field equations, so it is not a fixed constant from the biginning even if it has no preferred value 
at least at the classical level. It would be splendid if quantum effects could pick up a small and positive value among
arbitrary values via still unknown dynamical mechanism.   

We are now ready to present our idea. A natural question to ask ourselves is if it is possible to pay our attention 
to the first term in the action (\ref{EH action}), i.e., the Einstein-Hilbert term, and attempt to derive effectively 
the Newton constant as an integration constant from field equations by imposing a new constraint $\sqrt{-g} R = 1$. 

Of course, there is a big difference between unimodular gravity and our model. The unimodular constraint
$\sqrt{-g} = 1$ is a non-dynamical equation while the new constraint $\sqrt{-g} R = 1$ contains the second
derivative of the time variable so it is a dynamical one. In this sense, the constraint $\sqrt{-g} R = 1$
would control the dynamics of the model and provide us with a new perspective on the background geometry.

The structure of this article is the following: In Section 2, we present a simple model which accommodates
the curvature density preserving condition, and derive the equations of motion.
Moreover, we also construct a gravitational model respecting the full diffeomorphisms. 
In Section 3, we examine the cosmological implications coming from our model.
We conclude in Section 4.

\section{Gravitational models with curvature density preserving diffeomorphisms}

Let us start with a gravitational model where the Einstein-Hilbert term with the cosmological constant is
accompanied by the curvature density preserving constraint $\sqrt{-g} R = \varepsilon_0$: \footnote{Here 
for the slight generality, we have introduced a constant $\varepsilon_0$ instead of $\sqrt{-g} R = 1$.
Moreover, we have involved the Einstein-Hilbert plus the cosmological term for the generality. In this setting,
we can obtain induced gravity theory in the limit of the infinite Newton constant, $\kappa^2 = 8 \pi G
\rightarrow \infty$.}
\begin{eqnarray}
S = \int d^4 x  \left[ \frac{1}{2 \kappa^2} \sqrt{-g} \left( R - 2 \Lambda \right)
- \lambda(x) \left( \sqrt{-g} R - \varepsilon_0 \right) \right] + S_m,
\label{Action 1}
\end{eqnarray}
where $\lambda(x)$ is the Lagrange multiplier field enforcing the constraint $\sqrt{-g} R = \varepsilon_0$. 
Note that this action is not manifestly invariant under the full diffeomorphisms because of the presence 
of the term $\lambda(x) \varepsilon_0$. 

The Einstein equations, which come from variation with respect to the metric tensor $g^{\mu\nu}$, read
\begin{eqnarray}
\left[ \frac{1}{\kappa^2} - 2 \lambda(x) \right] G_{\mu\nu} + \frac{\Lambda}{\kappa^2} g_{\mu\nu}
+ 2 \left( \nabla_\mu \nabla_\nu \lambda - g_{\mu\nu} \nabla_\rho \nabla^\rho \lambda \right)
= T_{\mu\nu}.
\label{Einstein-eq 2}
\end{eqnarray}
Variation of the Lagrange multiplier field $\lambda(x)$, of course, produces our constraint $\sqrt{-g} R = \varepsilon_0$.

Operating $\nabla^\mu$ on the Einstein equations (\ref{Einstein-eq 2}) leads to
\begin{eqnarray}
- 2 \nabla^\mu \lambda \cdot G_{\mu\nu} + 2 \left( \nabla^2 \nabla_\nu \lambda - \nabla_\nu \nabla^2 \lambda \right)
= 0,
\label{Divergence-eq 1}
\end{eqnarray}
where the Bianchi identity and the conservation law of the energy-momentum tensor are used. Then, using the
identity $\nabla^2 \nabla_\nu \lambda = R_{\mu\nu} \nabla^\mu \lambda + \nabla_\nu \nabla^2 \lambda$, 
Eqs. (\ref{Divergence-eq 1}) can be cast to the form
\begin{eqnarray}
R \nabla_\mu \lambda = 0,
\label{Divergence-eq 2}
\end{eqnarray}
from which, assuming $R \neq 0$, we have $\lambda(x) = \lambda_0$ where $\lambda_0$ is an integration constant 
as in unimodular gravity.

Inserting $\lambda(x) = \lambda_0$ to the Einstein equations (\ref{Einstein-eq 2}), we arrive at the conventional
Einstein equations except a modified Newton constant $\kappa^\prime$ in front of the Einstein tensor 
which is defined as $\frac{1}{\kappa^{\prime 2}} = \frac{1}{\kappa^2} - 2 \lambda_0$:
\begin{eqnarray}
\frac{1}{\kappa^{\prime 2}} G_{\mu\nu} + \frac{\Lambda}{\kappa^2} g_{\mu\nu}
= T_{\mu\nu}.
\label{Einstein-eq 3}
\end{eqnarray}

At this stage, it is worthwhile to mention that in the limit of the infinite Newton constant, $\kappa^2 \rightarrow \infty$,
the gravitational model at hand reduces to an induced gravity \cite{Sakharov}-\cite{Oda2}. 
Actually, in this case, the Einstein-Hilbert plus the cosmological constant terms drop out of the classical action (\ref{Action 1}), 
and if we set $- 2 \lambda_0 = \frac{1}{\kappa^2}$, Eqs. (\ref{Einstein-eq 3}) become the Einstein equations 
without the cosmological constant
\begin{eqnarray}
G_{\mu\nu} = \kappa^2 T_{\mu\nu}.
\label{Einstein-eq 4}
\end{eqnarray}
Namely, in this specific situation, we start with only the matter action without the gravitational action, but our constraint
generates the gravitational dynamics in a natural way. The key point here is that the Lagrange multiplier field
$\lambda(x)$ takes any constant term as an integration constant, which is also a peculiar feature of unimodular gravity. 
 
In order to understand the constraint $\sqrt{-g} R = \varepsilon_0$, let us recall that general relativity is
invariant under diffeomorphisms whose infinitesimal forms are written as $\delta g_{\mu\nu} 
= \nabla_\mu \epsilon_\nu + \nabla_\nu \epsilon_\mu$ where $\epsilon_\mu$ are the infinitesimal local parameters.
Under diffeomorphisms, the constraint $\sqrt{-g} R = \varepsilon_0$ is transformed as
\begin{eqnarray}
\delta \left( \sqrt{-g} R - \varepsilon_0 \right) = \sqrt{-g} R \nabla_\mu \epsilon^\mu 
= \varepsilon_0 \nabla_\mu \epsilon^\mu.
\label{Diffeo 1}
\end{eqnarray}
Thus, as long as $\varepsilon_0 \neq 0$, the constraint $\sqrt{-g} R = \varepsilon_0$ can be interpreted as 
a gauge condition for one of diffeomorphisms. To put differently, as in unimodular gravity, this constraint
breaks diffeomorphisms down to the transverse diffeomorphisms (TDiff) which are defined as diffeomorphisms
satisfying $\nabla_\mu \epsilon^\mu = 0$. 

If one regards the above gravitational model as a gauge-fixed version of general relativity, together with 
the ghost term, the term involving the constraint in the action (\ref{Action 1}) can be combined into
a BRST-exact form, by which the interesting dynamical result makes no sense at least physically. 
Furthermore, as another unsatisfactory problem, when $R = 0$, Eq. (\ref{Divergence-eq 2}) does not give rise to
the result such that $\lambda(x)$ is a constant. It would be desirable if we could find a model which
holds even in case of $R = 0$ since black holes satisfy $R = 0$.

On the basis of the model constructed so far, however, it is easy to overcome such two problems at the same time
as follows: The key idea is to apply the Henneaux-Teitelboim method \cite{Henneaux} to our system, that is, the starting
action is defined as
\begin{eqnarray}
S = \int d^4 x  \left[ \frac{1}{2 \kappa^2} \sqrt{-g} \left( R - 2 \Lambda \right)
- \lambda(x) \left( \sqrt{-g} R - \varepsilon_0 \sqrt{-g} \nabla_\mu \tau^\mu \right) \right] + S_m,
\label{HT Action}
\end{eqnarray}
where $\tau^\mu$ is a vector field. Let us note that this action is manifestly invariant under the full 
diffeomorphisms.
  
Taking variation of the metric tensor leads to the Einstein equations:
\begin{eqnarray}
\left[ \frac{1}{\kappa^2} - 2 \lambda(x) \right] G_{\mu\nu} + \frac{\Lambda}{\kappa^2} g_{\mu\nu}
+ 2 \left( \nabla_\mu \nabla_\nu \lambda - g_{\mu\nu} \nabla^2 \lambda \right)
- 2 \varepsilon_0 \left[ \tau_{(\mu} \nabla_{\nu)} \lambda 
- \frac{1}{2} g_{\mu\nu} \tau_\rho \nabla^\rho \lambda \right]
= T_{\mu\nu},
\label{Einstein-eq 5}
\end{eqnarray}
where the round bracket indicates the symmetrization of indices of weight $\frac{1}{2}$.
Variation of the Lagrange multiplier field $\lambda(x)$, gives us the equation of motion:
\begin{eqnarray}
\sqrt{-g} R = \varepsilon_0 \sqrt{-g} \nabla_\mu \tau^\mu 
= \varepsilon_0 \partial_\mu \left( \sqrt{-g} \tau^\mu \right).
\label{Condition}
\end{eqnarray}
Finally, when one regards $\tau^\mu$ as a fundamental vector field, variation of $\tau^\mu$
leads to 
\begin{eqnarray}
\sqrt{-g} \varepsilon_0 \nabla_\mu \lambda = 0,
\label{Tau}
\end{eqnarray}
from which, the Lagrange multiplier field takes any constant value, $\lambda(x) = \lambda_0$.
Then, substituting $\lambda(x) = \lambda_0$ into Eqs. (\ref{Einstein-eq 5}), we have again
the standard Einstein equations (\ref{Einstein-eq 3}) except the Newton constant. 
As before, taking the limit of the infinite Newton constant and setting 
$- 2 \lambda_0 = \frac{1}{\kappa^2}$ yields to the Einstein equations without the cosmological 
constant (\ref{Einstein-eq 4}).

\section{Cosmological solutions}

In this section, first of all, we work with the final model with the full diffeomorphisms and 
consider the cosmological solutions to Eqs. (\ref{Einstein-eq 3}) and (\ref{Condition}) in the framework of 
the Friedmann-Robertson-Walker (FRW) universe with spacially flat metric since solutions in the other models
can be obtained in special cases from this model.

Now, taking the trace of Eqs. (\ref{Einstein-eq 3}), one obtains
\begin{eqnarray}
- \frac{1}{\kappa^{\prime 2}} R + \frac{4 \Lambda}{\kappa^2} = T.
\label{lambda-eq 2}
\end{eqnarray}
Next, eliminating $\Lambda$ in Eqs. (\ref{Einstein-eq 3}) via this relation gives us 
\begin{eqnarray}
\frac{1}{\kappa^{\prime 2}} \left( R_{\mu\nu} - \frac{1}{4} g_{\mu\nu} R \right) 
= T_{\mu\nu} - \frac{1}{4} g_{\mu\nu} T,
\label{Traceless Einstein-eq 2}
\end{eqnarray}
which are the traceless Einstein equations with the modified Newton constant.
Moreover,  Eqs. (\ref{Condition}) and (\ref{lambda-eq 2}) determine the vaule of the covariant
divergence of the vector $\tau^\mu$ 
\begin{eqnarray}
\nabla_\mu \tau^\mu = \frac{\kappa^{\prime 2}}{\varepsilon_0} \left( - T + 
\frac{4 \Lambda}{\kappa^2} \right).
\label{Cov-div}
\end{eqnarray}
Since this general model (\ref{HT Action}) includes an arbitrary vector field $\tau^\mu$, 
it is possible to find various types of cosmological solutions by adjusting the vector
field in an appropriate manner. 

Next we therefore wish to turn our attention to a more specific model
without the vector field.  Here we take account of the first gravitational model
where the equations of motion are given by the Einstein equations
\begin{eqnarray}
\frac{1}{\kappa^{\prime 2}} G_{\mu\nu} + \frac{\Lambda}{\kappa^2} g_{\mu\nu}
= T_{\mu\nu},
\label{Einstein-eq 6}
\end{eqnarray}
and the constraint equation
\begin{eqnarray}
\sqrt{-g} R = \varepsilon_0.
\label{Constraint eq}
\end{eqnarray}
The cosmological constant term can be always interpreted as the contribution of
vacuum energy to the Einstein equations, so let us henceforth include it in the
energy-momentum tensor of matter and set $\Lambda = 0$ in Eqs. (\ref{Einstein-eq 6}).

Since the constraint equation (\ref{Constraint eq}) includes second derivatives of time,
it is a dynamical equation which should be contrasted with the unimodular constraint
$\sqrt{-g} = v_0$, which is a non-dynamical one. In other words, our constraint (\ref{Constraint eq})
might restrict the whole class of classical equations satisfying the (modified) Einstein
equations (\ref{Einstein-eq 6}) to be its certain subgroup. Indeed, we will see shortly that 
only the universe filled with non-relativistic matter ("dust") is allowed as a classical solution. 

To do that, recalling that we have set $\Lambda = 0$, let us take the trace of Eqs. (\ref{Einstein-eq 6}),
\begin{eqnarray}
R = - \kappa^{\prime 2} T.
\label{Trace-eq}
\end{eqnarray}
Together this equation with Eq. (\ref{Constraint eq}), the trace part of the energy-momentum tensor
is described as
\begin{eqnarray}
T = - \frac{\varepsilon_0}{\kappa^{\prime 2}} \frac{1}{\sqrt{-g}} 
= - \frac{\varepsilon_0}{\kappa^{\prime 2}} a(t)^{-3},
\label{Trace-eq 2}
\end{eqnarray}
where we have worked with the FRW metric with spacially flat metric ($k = 0$)
\begin{eqnarray}
d s^2 = g_{\mu\nu} d x^\mu d x^\nu = - d t^2 + a(t)^2 ( d x^2 + d y^2 + d z^2 ),
\label{FRW metric}
\end{eqnarray}
with $a(t)$ being the scale factor. It is sufficient for our purposes to treat
matter as the perfect fluid $T^\mu \ _\nu = diag (-\rho, p, p, p)$ and
consider the equation of state $w = \frac{p}{\rho} = const.$ Then, the covariant
conservation law, $\nabla^\mu T_{\mu 0} = 0$ is solved to be $\rho(t) = \rho_0
a(t)^{-3 (1 + w)}$ where $\rho_0$ is an integration constant. Using these facts,
the trace part of the energy-momentum tensor takes the form
\begin{eqnarray}
T = - \rho + 3 p = \rho_0 ( -1 + 3w ) a(t)^{-3 (1 + w)}.
\label{Trace-eq 3}
\end{eqnarray}
By comparing this result (\ref{Trace-eq 3}) with Eq. (\ref{Trace-eq 2}), we find that
\begin{eqnarray}\\
w = 0, \quad \rho_0 = \frac{\varepsilon_0}{\kappa^{\prime 2}}.
\label{w}
\end{eqnarray}

Finally, using Eq. (\ref{w}), the Einstein equations (\ref{Einstein-eq 6}) are solved and the scale factor
is completely determined to be
\begin{eqnarray}\\
a(t) = \left( \frac{3}{4} \varepsilon_0 \right)^{\frac{1}{3}} \left( t - t_0 \right)^{\frac{2}{3}},
\label{Scale factor}
\end{eqnarray}
where $t_0$ is some constant. This solution describes the decelerating universe filled with
non-relativistic matter whose equation of state is $w = 0$.

\section{Conclusion}

In this article, along the similar line of the argument to unimodular gravity,
we have constructed gravitational models which have either the transverse 
diffeomorphisms or the full diffeomorphisms. 
In the limit of the infinite Newton constant, these models reduce to those of induced gravity. 
Moreover, we have investigated classical solutions which satisfy the equations of motion.

One of interesting features in unimodular gravity is that the cosmological constant appears
as an integration constant unrelated to any parameters in the action. Usually, this 
feature has been utilized in order to solve the well-known cosmological constant problem.
However, in this article, we have made use of this feature to show that the Newton constant appears as 
an integration constant from the induced gravity action.  In modern times, we are tempted to
consider that quantum gravity is not only an emergent phenomenon but also independent of background
metric. In such a viewpoint, the present work might shed some light on quantum gravity.

\begin{flushleft}
{\bf Acknowledgements}
\end{flushleft}
This work is supported in part by the Grant-in-Aid for Scientific 
Research (C) No. 25400262 from the Japan Ministry of Education, Culture, 
Sports, Science and Technology.


\end{document}